\documentclass[a4paper,reqno]{amsart}
\usepackage[text={5.2in,8in},centering]{geometry}
\usepackage[backrefs]{amsrefs}
\usepackage{amssymb}
\usepackage{enumerate}
\usepackage{graphicx}
\usepackage{color}
\definecolor{trustcolor}{rgb}{0.71,0.14,0.07}

\numberwithin{equation}{section}
\theoremstyle{plain}

\newtheorem{theorem}{Theorem}

\newtheorem{prop}[theorem]{Proposition}
\newtheorem{cor}{Corollary}[section]
\newtheorem{lemma}{Lemma}[section]

\theoremstyle{remark}
\newtheorem{remark}{Remark}[section]

\newtheorem*{quest*}{Question}
\newtheorem*{remark*}{Remark}
\theoremstyle{remark}

\theoremstyle{definition}
\newtheorem{definition}{Definition}[section]
\newtheorem*{definition*}{Definition}
\newtheorem*{notation*}{Notation}
\newtheorem*{notations*}{Notations}
\providecommand{\B}{\mathbf}

\providecommand{\C}{\mathcal}
\providecommand{\D}{\mathbb}

\providecommand{\R}{\mathrm}

\newcommand{\ee}{\mathrm{e}}

\DeclareMathOperator{\boxx}{B}

\def\boxxk{\boxx_{L_k}}
\def\boxxkone{\boxx_{L_{k+1}}}

\def\boxxkfour{\boxx_{L^{4}_{k}}}

\DeclareMathOperator{\dist}{d}

\DeclareMathOperator{\tr}{tr\,}
\DeclareMathOperator{\card}{card}

\DeclareMathOperator*{\essup}{ess\,sup}

\DeclareMathOperator{\loc}{loc}

\DeclareMathOperator{\one}{\mathbf{1}}

\DeclareMathOperator{\supp}{{\rm supp}}

\DeclareMathOperator{\diam}{{\rm diam}}

\def\loc{$m\mathcal{-L}${\textit{oc}}}

\def\thomnloc{$(m,\om,\th)\mathcal{-N}\mathcal{L}${\textit{oc}}}
\def\thomloc{$(m,\om,\th)-\mathcal{L}${\textit{oc}}}

\def\ual{\B{\alpha}}

\def\DC{\D{C}}

\def\DE{\D{E}}
\def\DN{\D{N}}
\def\DP{\D{P}}

\def\DR{\D{R}}
\def\DT{\D{T}}

\def\DZ{\D{Z}}

\def\cA{\C{A}}

\def\cC{\C{C}}
\def\cD{\C{D}}

\def\cM{\C{M}}
\def\cN{\C{N}}

\def\cR{\C{R}}
\def\cS{\C{S}}
\def\cT{\C{T}}
\def\cX{\C{X}}

\def\cW{\C{W}}

\def\hu{\hat{u}}

\def\Const{{\R{Const\, }}}
\def\Cdprime {{C''}}

\def\om{\omega}
\def\Om{\Omega}

\def\Lam{\Lambda}
\def\lam{\lambda}

\def\th{\theta}
\def\Th{\Theta}

\def\ffi{\varphi}

\def\fB{\mathfrak{B}}
\def\fBtj{\widetilde{\mathfrak{B}}_{<j}}

\def\fF{\mathfrak{F}}
\def\fR{\mathfrak{R}}
\def\fS{\mathfrak{S}}

\def\pt{\partial}
\def\half{\frac{1}{2}}

\def\truc#1#2#3{\smash{\mathop{\,\, #1 \,\, }\limits^{#2}_{#3}}}

\def\tto#1{\smash{\mathop{\,\,\,\, \longrightarrow \,\,\,\, }\limits_{#1}}}

\def\myset#1{{\left\{\,#1\,\right\}}}
\def\exist{\exists\,}

\def\Thinf{{\Th_{\infty}}}

\def\USR{{$\mathbf{(USR)}$}}
\def\DIV{{$\mathbf{(DIV)}$}}
\def\LVB{{$\mathbf{(LVB)}$}}

\def\Sparsek{{$\mathbf{(Sparse(k))}$}}
\def\Sparsez{{$\mathbf{(Sparse(0))}$}}
\def\Sparsekone{{$\mathbf{(Sparse(k+1))}$}}

\def\diy{\displaystyle}

\def\ba{\begin{array}{l}}
\def\ea{\end{array}}
\def\bal{\begin{aligned}}
\def\eal{\end{aligned}}
\def\be{\begin{equation}}
\def\ee{\end{equation}}

\def\Sigomth#1{{\Sigma_{#1}}}

\def\pr#1{{\DP\left\{\,#1\,\right\}}}
\def\prpmu#1{{\DP^{\Om\times\Th}\left\{\,#1\,\right\}}}

\def\esmmu#1{{\DE^{(\th)}\left[\,#1\, \right]}}
\def\esmommu#1{{\DE^{\Om\times\Th}\left[\,#1\, \right]}}
\def\prmu#1{{\mu\left\{\,#1\,\right\}}}
\def\prmuj#1{{\mu_{j-1}\left\{\,#1\,\right\}}}

\def\prth#1{{\mu\left\{\,#1\,\right\}}}

\def\mymax#1{{ \truc{\max} {} {#1}}}

\def\qube#1#2{{Q_{#1}(#2)}}

\def\tabhigh#1{{\buildrel _{} \over {#1}}}

\begin{document}
\title[Anderson  localization for generic deterministic operators]
{Anderson localization  \\ for generic deterministic operators }


\author[V. Chulaevsky]{Victor Chulaevsky}


\address{D\'{e}partement de Math\'{e}matiques\\
Universit\'{e} de Reims, Moulin de la Housse, B.P. 1039\\
51687 Reims Cedex 2, France\\
E-mail: victor.tchoulaevski@univ-reims.fr}

\date{}
\begin{abstract}
We consider a  class of ensembles of lattice Schr\"{o}dinger operators with deterministic random potentials, including quasi-periodic potentials with Diophantine frequencies, depending upon an infinite number of parameters in an auxiliary measurable space. Using a variant of the Multi-Scale Analysis, we prove Anderson localization for generic ensembles in the strong disorder regime and establish an analog of Minami-type bounds for spectral spacings.
\end{abstract}
\maketitle

\section{Introduction. Formulation of the results.}\label{intro}

In this paper, we study spectral properties of finite-difference operators, usually called lattice Schr\"{o}dinger operators (LSO), of the form
$$
(H(\om,\th) f)(x)
= \sum_{y:\, \|y-x\|_1=1}f(y) + gv(T^x \om,\th) f(x),\; x,y\in \DZ^d,
$$
where $v: \Omega\times\Theta\to\DR$ is a measurable  function on the direct product of the probability space $(\Omega=\DT^\nu, \fF, \DP)$, endowed with the  normalized Haar measure $\DP$, and an auxiliary probability  space $(\Theta,\fB,\mu)$. $T:\DZ^d\times\Om\to\Om$ is an ergodic dynamical system with discrete time $\DZ^d$, $d\geq 1$. Here $\|x\|_1=\sum_i|x_i|$. The function $v$ will be referred to as the \emph{hull} of the ergodic potential $V$. In an earlier work \cite{C09}, we considered a particular case where the hull $v$ was discontinuous on the torus, for a.e. $\th\in\Th$. In fact, the arguments of \cite{C09} were based on a particular construction which \emph{required} $v$ to be discontinuous. In the present paper, we show that methods of \cite{C09} can be adapted to the hulls of class $\cC^M(\DT^\nu)$ for any given $M\ge 1$. The probability space $\Th$ plays the role of a parameter space with  measure $\mu$ constructed in such a way that $\mu$-a.e. hull
$\om\mapsto v(\om,\th)$ is suitable for the Multi-Scale Analysis (MSA), so that the exponential spectral localization can be established for $\mu$-a.e. ergodic ensemble of operators $H(\cdot;\th)$ (with $g$ large enough).

Recall that in the works by Sinai \cite{Sin87} and by Fr\"{o}hlich et al. \cite{FSW90} Anderson localization was proven for the one-dimensional LSO with quasi-periodic potential of the form
$V(x;\om) = v(\om + x\alpha)$, $\om\in\DT^1$, where the hull $v:\DT^1\to\DR$ was assumed of class $\cC^2(\DT^1)$ with exactly two critical points, both non-degenerate. Later, it became clear that in order to extend these  techniques to more general hull functions and multi-dimensional phase spaces, e.g., $\Om = \DT^\nu$ with $\nu>1$, it would be necessary to exclude an infinite number of `degeneracies' which cannot be described explicitly; cf. \cite{CS89}.  Here we show that necessary regularizations, required in the course of the  MSA procedure, can be performed  with the help of relatively simple probabilistic techniques.

Bourgain, Goldstein and Schlag developed earlier a different approach; see, e.g., \cite{BG00,BS00,BGS01}. Their method is based on the assumption of analyticity of the hull  $v:\DT^\nu\to\DR$.

Auxiliary parameters allowing to avoid "small denominators" in the MSA procedure can be introduced in many ways. For example, it is quite natural to consider a series with `random' coefficients $a_n(\th)$ relative to a probability space $(\Th,\fB,\mu)$
$$
v(\om,\th) = \sum_{n\in\DZ} a_n(\th) \ffi_n(\om)
$$
and wavelet-like functions $\ffi_n$. It turns out, however, that the orthogonality properties are of little importance here. Earlier, we proposed in \cite{C01,C07} a class of parametric families of deterministic potentials which we called "randelette expansions"; cf. Section \ref{sec:GE.examples}.

An important class of examples is obtained by taking an ergodic\footnote{Although ergodicity \emph{per se} is not required  for the proof of localization, it follows from \USR\, for rotations of the torus.}
action of the group $\DZ^d$ on  the torus $\DT^\nu$, $\nu\ge 1$, generated by quasi-periodic shifts
$$
T_{\ual_j}:\, \om  \mapsto  \om + \ual_j, \quad  \ual_j\in\DT^\nu.
$$
Recently, Chan \cite{Chan07} proved the  Anderson localization for single-frequency quasi-periodic operators with the hull $v$ of class $C^3(\DT^1)$, using a parameter exclusion technique which is different from presented in this paper.

Our main requirement for the dynamical system is the condition \eqref{eq:USR} of `uniformly slow' returns of any trajectory $\myset{T^x\om, x\in \DZ^d}$ toward its starting point $\om\in\Om$.

\medskip

\textbf{The main result } of the present paper, Theorem \ref{thm:main.SL}, is formulated in Section \ref{ssec:main.result}.

An interesting "by-product" of our approach is a Minami-type bound; cf. Section \ref{sec:Minami}.

\subsection{Requirements for the dynamical system}\label{SubSecUSR}

We assume that the underlying dynamical system $T$ on the phase space $\Om$, endowed with a distance $\dist_\Om(\cdot\,,\cdot)$, satisfies the following condition of \emph{uniformly slow returns}:

\smallskip
\USR: $\exist\, A,C\in(0,\infty)\;\; \;\forall\, \om\in\Om \;\;\forall\, x,y\in\DZ^d, x\ne y$
\be\label{eq:USR}
\begin{array}{lc}
\quad \dist_{\Om}(T^x \om, T^y \om) \ge 4C \|x - y \|^{-A}.
\end{array}
\ee
\smallskip
In this paper, we consider  the case where $\Om = \DT^\nu$, $\nu\ge 1$, and it is  convenient to define the distance $\dist_\Om(\om', \om'') \equiv \dist_{\DT^\nu}(\om', \om'')$ as follows:
$$
\dist_{\DT^\nu}\big((\om'_1, \ldots, \om'_\nu), (\om''_1, \ldots, \om''_\nu)\big)
= \max_{1 \le i \le \nu} \dist_{\DT^1}(\om'_i, \om''_i),
$$
where $\dist_{\DT^1}$ is the conventional distance on the unit circle $\DT^1$.  The reason for the choice of the phase space $\Om=\DT^\nu$ is that many parametric families of ensembles of potentials $V(x;\om,\th)$ and various dynamical systems can be made  explicit in this case; in fact, the torus can be replaced by a compact Riemannian manifold of class $\cC^M$, $M\ge 1$.
For the rotations of the torus $\DT^\nu$,
$$
T^x \om = \om + x_1 \alpha_1 + \cdots + x_d \alpha_d,
\; x \in\DZ^d, \; \alpha_j\in \DT^\nu, \, 1 \le j \le d,
$$
the USR property reads as a Diophantine condition for the frequency vectors $\ual_j$.
Recall that, owing to a well-known result by Gordon \cite{G76}, a quasi-periodic operator with irrational frequency abnormally fast approximated by rational numbers may have no decaying  solution to the problem $H\psi = E\psi$ (hence, no $\ell^2$-eigenfunction).

We also assume a polynomial bound on the rate of local divergence of trajectories (fulfilled for rotations of the torus, as well as for  skew shifts):

\smallskip
\DIV: $\exist\, A',C' \in(0,\infty)\; \;\forall\, \om,\om'\in\Om \;\;\forall\, x\in\DZ^d $
\be\label{eq:DIV}
\begin{array}{lc}
\quad \dist_{\Om}(T^x \om, T^x \om') \le C'  \|x \|^{A'} \dist_\Om(\om, \om').
\end{array}
\ee

\subsection{Geometrical notions and constructions}
$\,$

We will consider lattice cubes
$
\boxx_L(u) = \myset{x\in\DZ^d:\; \|x-u\| \le L},
$
for which we define the internal boundary
$
\pt^{-} \boxx_L(u) = \myset{x:\; \|x-u\| = L},
$
external boundary composed of nearest neighbors $y\in\DZ^d\setminus\boxx_L(u)$ of
$\pt^{-} \boxx_L(u)$,
and boundary $\pt\boxx_L(u)$ composed of nearest-neighbor pairs $(x,y)$, with
$x\in\pt^{-} \boxx_L(u)$ and $y\in\pt^{+} \boxx_L(u)$.
Here and below, we use the max-norm for vectors $x\in\DR^d$:
$
\diy\|x\|  := \max_{1\le i \le d} |x_i|,
$
so that cubes $\boxx_L(u)$ are actually balls of radius $L$ centered at $u$. For this reason, we will often refer to the "radius" ($=L$) of a cube $\boxx_L(u)$.
We will work with restrictions of the operator $H(\om,\th)$ to lattice cubes $\boxxk(u)$ with Dirichlet boundary conditions on $\pt^{+} \boxxk(u)$, where $L_0>2$ is a sufficiently large integer and $L_k$, $k\ge 1$, are defined by the recursion
\be\label{eq:def.L.k}
L_k = [L_{k-1}]^\alpha + 1, \; \alpha=3/2.
\ee
Next, define a sequence of positive real numbers
\be
 \delta_k = g^{-a} e^{-4 L_k^{b}}, \, k\ge 0; \quad a = 1/2, \;b = 1/2.
\ee
It is convenient to introduce also the scale length $L_{-1} = 0$ (cf., e.g., Definition \ref{def:E.CNR}). Cubes $\boxx_{0}(u)$ are
single-point sets: $\{x:\, \|x-u\| \le 0\} \equiv \{u\}$. The
restriction of operator $H(\om,\th)$ on a $\boxx_{L_{-1}}(u)$ is the operator of
multiplication by $V(u;\om,\th)$.

The spectrum of operator $H_{\boxx}(\om,\th)$ in a cube $\boxx$ will be denoted by
$\Sigomth{\om,\th}(\boxx)$.

We will also use cubes $\qube{r}{\om'}\subset\DT^\nu$ of the form
$$
\qube{r}{\om'} = \{\om\in\Om:\; \dist_{\DT^\nu}(\om, \om')\le r \}, \; r>0.
$$
We always assume that $g>0$.

\subsection{Local Variation Bound}
\label{ssec:LVB}

The random field $v:\Om\times\Th\to\DR$ on $\Om$  relative to the auxiliary probability space $(\Th,\fB,\mu)$ is assumed to fulfill the following condition:

\smallskip
\LVB:
\textit{
There exists a family of sigma-algebras $\fB(L,x)\subset\fB$, labeled by non-negative integers $L$ and lattice points $x$, such that for all $L$ and $x$
\begin{enumerate}[\rm (i)]
  \item all random variables $v(T^y \om,\th)$ with $y\in\boxx_{L^{4}}(x)\setminus\{x\}$ are
  $\fF\times\fB(L,x)$-measurable;
  \item  the random variable $(\om,\th)\mapsto v(T^x \om,\th)$ on $\Om\times\Th$ admits a bounded conditional probability density $p_{v,x}(\, \cdot\, |\fF\times\fB(L,x) )$:
\be\label{eq:LVB}
   \| p_{v,x}(\,\cdot\, |\fF\times\fB(L,x) ) \|_\infty \le \Cdprime   L^B, \;\; \Cdprime\in(0,+\infty).
\ee
\end{enumerate}
}

In other words, for any lattice cube $\boxx_L(u)$, even with the phase point $\om\in\Om$
and all values $\{v(y), y\in\boxx_L(x), y\ne x\}$ fixed, there is `enough parametric freedom' in the potential value $V(x,\om,\th)$ to guarantee absolute continuity of its (conditional) probability distribution.

To clarify the nature of the sigma-algebras $\fB(L,x)$, which are constructed explicitly in Section \ref{sec:GE.examples} for a particular class of hulls $v:\Om\times\Th\to\DR$, note that there actually exist sigma-algebras $\fB_L\subset\fB$, labeled only by scales $L>0$, such that conditional on $\fB_L$ and on $\fF$ (i.e., with $\om$ fixed) all values of the potential
$\{V(x,\om,\th), x\in\boxx_L(u)\}$ in \emph{any} cube of radius $L$ are conditionally independent and admit individual (non-identical) conditional probability densities $p_x(\cdot\,|\fF\times\fB_L)$ uniformly bounded by $O(L^B)$. This property gives rise to Wegner-type and Minami-type bounds in finite cubes, although these bounds deteriorate as the size of the cube grows. Furthermore, the exponent $B$ of the power-law growth of $\| p_x(\cdot\,|\fF\times\fB_L) \|_\infty$ depends upon the rate of returns figuring in the assumption \USR\, and becomes higher when the exponent $A$ in \eqref{eq:USR} grows.

Following \cite{C01,C07}, we will call a random field satisfying \LVB\, a regular \emph{grand ensemble}. Examples of such ensembles are given in Section \ref{sec:GE.examples}. It is worth mentioning that for any $M\ge 1$ there exist quite natural "grand ensembles" with \emph{all} samples $v(\om,\cdot)$ of class $\cC^M(\Om)$. Moreover, there exist regular grand ensembles with  \emph{discontinuous} samples for which one can prove  Anderson localization in strong disorder regime; cf. \cite{C09}. On the other hand, \LVB\, says that the local interpolation problem for the field $v(\om,\cdot)$ on $\Om$, relative to $(\Th,\fB,\mu)$, must not admit an exact solution. This explains why our approach does not allow to treat analytic hulls. So, from several points of view, it is complementary to the approach developed by Bourgain, Goldstein and Schlag.

Note also that if $v(\om,\th)$ is a regular grand ensemble of class $\cC^M$ and $W\in\cC^M(\Om)$ is an arbitrary "background potential", then the operators $H = \Delta + W(\om) + gV(\om,\th)$ feature Anderson localization for sufficiently large $g$; in fact, the background potential $W:\Om\to\DR$ would play almost no role in our analysis.

In this paper, we make one more assumption: a uniform boundedness of the gradient of the sample hull functions $\om\mapsto v(\om, \th)$:
\be\label{eq:gradient.bounded}
\exists\, \tilde C<\infty\; \forall\, \th\in\Th
\quad \| v(\cdot, \th) \|_{\cC^1(\DT^\nu)} \le \tilde C.
\ee
The construction of the grand ensemble with the help of "randelette expansions" in Section \ref{sec:GE.examples} guarantees \eqref{eq:gradient.bounded}. In a more general context, one would need to exclude samples with excessively large gradients. Clearly, if
$\prmu{ \| v(\cdot, \th) \|_{\cC^1(\DT^\nu)} < \infty}=1$, then the set of samples with large gradients must have a small $\mu$-measure.

It is readily seen that for the random variables $gv(x;\om,\th)$, Eqn \eqref{eq:LVB} gives rise to
\be\label{eq:LVB.g}
\| p_{gv,x}\|_\infty \le \Cdprime   g^{-1}L^{B}.
\ee

We will denote by $\prpmu{\cdot}$ the product measure $\DP\times\mu$ on $\Om\times\Th$ and by
$\esmommu{\cdot}$ the respective expectation.

\subsection{Main result}
\label{ssec:main.result}

\begin{theorem}\label{thm:main.SL}
Consider a family of lattice Schr\"{o}dinger operators in $\ell^2(\DZ^d)$,
$$
H(\om,\th) = \Delta + gV(x;\om,\th)
$$
where  $V(x;\om,\th) = v(T^x\om,\th)$. Suppose that the dynamical system $T:\DZ^d\times\Om\to\Om$  and the function $v:\Om\times\Th\to\DR$ satisfy the conditions \USR, \LVB\, and \DIV\, for some $A,B,C,C',\Cdprime  \in(0,\infty)$.
For sufficiently large $g\ge  g_0(A,B,C,C',\Cdprime  )$,  there exists a subset
$\Thinf(g) \subset \Theta$ of measure $\prth{\Thinf(g)} \ge 1 - c(A,B,C,C',\Cdprime)\, g^{-1/2}$ with the following property: if $\th\in\Thinf$, then for $\DP$-a.e. $\om\in\Om$
the operator $H(\om,\th)$ has pure point spectrum, and for every eigenfunction $\psi_j\in\ell^2(\DZ^d)$ there exist
$\hu_j\in\DZ^d$ and  $L\in\DN$ such that for all $x$ with $\|x-\hu_j\|\ge L$
$$
|\psi_j(x)| \le e^{-m\|x-\hu_j\|}
$$
with
$
m = m(g, A, B,C,C',\Cdprime  )> c_1(A,B,C,C',\Cdprime  ) \ln g > 0.
$
\end{theorem}

\noindent
\textbf{\textit{An informal outline of the proof.}}
\begin{enumerate}
  \item We use the general strategy of the MSA which requires, at each scale $L_k = L_0^{\alpha^k}$, $\alpha>1$,  two kinds of estimates:
\begin{itemize}
  \item an eigenvalue concentration (EVC) bound for the probability of having two disjoint cubes of radius $L_k$ inside a larger cube $\boxx_{L_{k+1}}(w)$ with spectra  abnormally close to each other, at some energy $E$ ("$E$-resonant" cubes);
  \item a  bound for the probability to have at least
  $\nu+2 \equiv \rm{dim } \Om + 2$ cubes of radius $L_k$ inside a larger cube $\boxx_{L_{k+1}}(w)$ in which the decay of the matrix elements of the resolvent is insufficient ("singular" cubes).
\end{itemize}

  \item The EVC bound in our case is proven in two ways:
\begin{itemize}
  \item  for a large set of parameters $\th\in\Th$ and \textbf{for any} phase point $\om\in\Om$, the maximal number of simultaneously "resonant" cubes in $\boxx_{L_{k+1}}(w)$ does not exceed $\nu+1$;
  \item the $\DP$-probability to have at least two "resonant" cubes in $\boxx_{L_{k+1}}(w)$ is small (sufficient for the purposes of the MSA).
\smallskip

      \emph{The first property rules out -- in a deterministic way -- an accumulation of resonant cubes, which is inevitable (albeit unlikely) in the case of random (e.g., IID) potentials. The second is necessary for the MSA; it is simpler to prove than the first one. Both of them are proven without scale induction.}
\smallskip

  It is worth noticing that at the initial scale $L_0$, under the assumption of large disorder, "non-resonant" cubes are also "non-singular". Therefore, the first property rules out an accumulation of "singular" cubes at least at the initial scale. This "sparseness" property is then to be proven inductively at all scales; see the proof of Lemma \ref{lem:main.ind}.
\end{itemize}
\smallskip

  \item Finally, we modify the traditional MSA tactics, which delays the analysis of the eigenfunction decay until the last stage where the finite-volume bounds of Green functions are established at any scale. Namely, we make use of the Geometric Resolvent Inequality (GRI) for eigenfunctions at scale $L_k$ and derive from the "sparseness" of "singular" cubes and EVC estimates a lower bound on the probability of having \emph{all} eigenfunctions exponentially decaying in a cube $\boxx_{L_{k}}(u)$. The spectral localization in $\DZ^d$ is proven in the usual way.

\end{enumerate}

\section{Scale induction }

\subsection{Resonances and tunneling}

\begin{definition}\label{def:E.CNR}
Given a real number $E$, a cube $\boxx_{L_{k}}(u)$, $ k\ge 0$, is called
\begin{itemize}
  \item $(E,\om,\th)$-non-resonant ($(E,\om,\th)$-NR) if
\be
\dist(\Sigomth{\om,\th}(\boxx_{L_{k}}(u)), \, E ) \le
g \delta_k \equiv g^{1-a}  e^{-4 L_k^{b}},
\ee
  and $(E,\om,\th)$-resonant ($(E,\om,\th)$-R), otherwise;
  \item $(E,\om,\th)$-completely non-resonant ($(E,\om,\th)$-CNR) if it does not contain  $(E,\om,\th)$-resonant cubes  of radius $\ell\ge L_{k-1}$ (including itself); otherwise, it is called $(E,\om,\th)$-partially resonant ($(E,\om,\th)$-PR);
  \item $(\om,\th)$-tunneling ($(\om,\th)$-T) if for some $E\in \DR$ it contains two disjoint $(E,\om,\th)$-PR cubes of radius $\ge L_{k}^{1/4}$, and $(\om,\th)$-non-tunneling ($(\om,\th)$-NT), otherwise;
  \item $(\om,\th)$-multi-resonant ($(\om,\th)$-MR) if for some $E\in \DR$ it contains at least\footnote{Recall that $\nu$ is the dimension of the phase space $\Om = \DT^\nu$.}
   $\nu+2$ disjoint $(E,\om,\th)$-PR cubes of radius $L_{k-1}$; otherwise, it is called $(\om,\th)$-NMR.
\end{itemize}
\end{definition}

\begin{definition}\label{def:E.m.NS}
Given  real numbers $E$ and $m>0$, a cube $\boxx_{L_{k}}(u)$ is called
\begin{itemize}
  \item $(E,m,\om,\th)$-non-singular ($(E,m,\om,\th)$-NS) if
\be\label{eq:def.NS}
\max_{ \|x - u \| \le L_{k-1}} \sum_{(y,y')\in\pt \boxx_{L_{k}}(u)}
|G_{\boxx_{L_{k}}(u)}(x,y;E;\om,\th)|
\le e^{- \gamma(m,L_k) L_k },
\ee
where
\be
\gamma(m,L) := m(1 + L^{-1/8}) >m,
\ee
  otherwise, it is called $(E,m,\om,\th)$-singular ($(E,m,\om,\th)$-S);
  \item $(m,\om,\th)$-bad, if for some $E\in \DR$,  if it contains at least $\nu+2$ disjoint cubes of radius $L_{k-1}$ which are $(E,m,\om,\th)$-S, and $(m,\om,\th)$-good, otherwise;
\end{itemize}
\end{definition}

We will need the following two analogs of the well-known Wegner bound:

\begin{lemma}[Wegner-type bounds]
\label{lem:Wegner}

For any $k\ge 0$ and any $u\in\DZ^d$,
\begin{enumerate}[\rm (A)]
  \item for any $E\in\DR$  and $s\in(0,1]$
\be\label{eq:Wegner.1}
\prpmu{(\om,\th):\,    \dist( \Sigma_{\om,\th}(\boxx_{L_k}(u)), E) \le s}
\le C_0 L_k^{B+d} g^{-1}\, s \,;
\ee
  \item for any $s\in(0,1]$ and  $k\ge 1$
\be\label{eq:Wegner.2}
\prmu{\th\,|
\pr{\om:\, \boxx_{L_k}(u) \text{ is $(m,\om,\th)$\rm{-T} } } \ge s}
\le
C_0   L_k^{B+3d}   \, \delta_k \,s^{-1}.
\ee
In particular, for $L_0$ large enough, any $k\ge 1$ and   $u\in\DZ^d$
\be\label{eq:Wegner.3}
\prmu{\th\,|\,
\pr{\om:\, \boxx_{L_k}(u) \text{ is $(\om,\th)$\rm{-T} } } \ge \delta_{k-1}^{1/2} }
\le  g^{-a/2} e^{-L_{k-1}^{b}}.
\ee
\end{enumerate}

\end{lemma}

We will also need a global bound on the number of resonant cubes.

\begin{lemma} For $L_0$ and $g$ large enough, any $k\ge 1$ and   $u\in\DZ^d$
\label{lem:main.NMR}
\be\label{eq:lem.main.NMR}
\prmu{\th|\; \exists\, \om\in\Om:\; \text{ $\boxx_{L_{k+1}}(u)$ is $(\om,\th)$\emph{-MR}} }
 < C_1 \; L^{J_\nu(3d + B + 1)} \, g^{-a/2} e^{-2L_k^{b}}.
\ee
\end{lemma}

In the situation where the grand ensemble $v(\om,\th)$ is given by a regular randelette expansion (cf. Section \ref{sec:GE.examples}), the Wegner-type bound was proven in our earlier work \cite{C01}. We prove Lemma \ref{lem:Wegner} in Appendix, using the properties \LVB\, and \USR.

The proof of Lemma \ref{lem:main.NMR} is  more involved; it is based on a probabilistic bound (cf. Lemma \ref{lem:small.prob.qube} below), combined with an analytic argument (cf. Lemma \ref{lem:small.var.partition}).

\subsection{Collections of resonant cubes}
\label{ssec:R.boxes.scale.Lk}

The results of this section do not use the scale induction and apply to any scale $L_k$, starting from $k=0$.

\begin{lemma}\label{lem:small.prob.qube}

Fix a point $\om'\in\DT^\nu$, set $J_\nu=\nu+2$  and consider the event  of the form
$$
\bal
\cR(k,\om') = \;&\big\{\th:\, \text{$\exist\,$ disjoint cubes
$\boxx_{R_j}(v^{(j)})\subset \boxx_{L_k}(u^{(j)})\subset\boxxkfour(0)$,
$1\le j \le J_\nu,$}
\\
&\;\,\text{with  $R_j\in[L_{k-1}, L_k]$ and such that for $j=2, \ldots, J_\nu$,}
\\
&\;\dist\big(\Sigma_{\om',\th}(\boxx_{R_j}(v^{(j)})), \Sigma_{\om',\th}(\boxx_{R_1}(v^{(1)})) \big)
\le  4 g \delta_k \big\} \subset \Th.
\eal
$$
Then
\be\label{eq:lem.small.prob.qube}
\bal
\prmu{ \cR(k,\om')}
\le C_1 \, L_k^{J_\nu(3d + B+1)} \, \delta_k^{J_\nu-1}.
\eal
\ee
\end{lemma}
\proof
Fix  an arbitrary cubes $\boxx_{R_i}(v^{(j)})\subset \boxx_{L_k}(u^{(j)})$ as in $\cR(k,\om')$ and consider sigma-algebras $\fB_j :=\fB(L_k^4, u^{(j)})$, $1\le j \le J_\nu$, figuring in the condition \LVB.

Observe that, by construction, for each $j\ge 2$,  all eigenvalues of operators $H_{\boxx_{R_i}(u^{(i)})}$ with $i<j$ are $\fF\times\fB(L_k^4,u^{(j)})$-measurable, since
$\boxx_{L_k}(u^{(i)})\subset\boxxkfour(0)\setminus\boxx_{L_k}(u^{(i)})$. So, conditioning on $\fF\times\fB(L_k^4,u^{(j)})$ fixes $\om$ and spectra of $H_{\boxx_{R_i}(u^{(i)})}$,
$i<j$.
With $\om=\om'$ fixed, all spectra become functions of $\th\in\Th$.
Let
$$
\bal
\boxx_i &= \boxx_{R_i}(u^{(i)}), & 1 \le i \le J_\nu, \\
\Sigma_i &=\Sigma_i(\om';\th) = \Sigomth{\om',\th}(\boxx_i),
& 1 \le i \le J_\nu,\\
\cD_i &= \cD_i(\om') = \big\{ \th:\, \dist(\Sigma_1, \Sigma_i)\le 4 g \delta_k \big\},
     & 2 \le i \le J_\nu, \\
\eal
$$
and denote by $\fBtj$ the sigma-algebra  generated by  sigma-algebras
$\{\fB_i, \; 1 \le i < j\}$.
Then for every $2 \le j \le J_\nu$, we can write, using the inequality $R_i \le L_k$:
$$
\prmu{ \cD_j \,\big|\, \fBtj }
\le (2L_k + 1)^{2d} \;
\max_{\lam^{(j)}_{r}\in \Sigma_j, \lam^{(1)}_{s}\in\Sigma_1}
\prmu{ |\lam^{(j)}_{r} - \lam^{(1)}_{s}|\le 4 g \delta_k \,\big|\, \fBtj }.
$$
As was noticed,  for fixed $\om'$, eigenvalues $\th\mapsto\lam^{(1)}_{s}(\om',\th)$  are $\fBtj$-measurable; we work with these random variables $\lam^{(1)}_{s}(\om',\cdot)$ and denote by
$\prmuj{\cdot}$ the conditional measure
$\mu\left\{ \cdot \,|\,   \fBtj\right\}$.
It suffices to bound the probability
$
\prmuj{ |\lam^{(j)}_{r} - E|\le 4 g \delta_k }
$
for any fixed $E\in\DR$. This can be done with the help of the conventional Wegner bound. Indeed, consider operator $H_{\boxx_j}$ in the cube
$\boxx_j$. For the Wegner  bound to apply, it suffices that for each point
$x\in\boxx_j$ the random variable $gV(x;\om,\th)$ admit a bounded  probability density,
conditional\footnote{As was pointed out in Section \ref{ssec:LVB}, for grand ensembles constructed in Section \ref{sec:GE.examples}, there actually exist sigma-algebras $\fB_L\subset\fB$ such that, conditional on $\fF\times\fB_L$, all values
$v(y;\om,\th)$ with $y\in\boxx_{L}(x)$ and \emph{any} $x$ are independent and admit individual
conditional densities bounded by $O(L^B)$.  }
on all other values of the potential
$\{ gV(y;\om,\th), y \in \boxx_j \setminus \{x\} \}$.

Recall that owing  to the assumption (\LVB, (i)), all values $V(y;\om,\th)$ with
$x\ne y\in \boxx_{L_k}(u^{(j)})$ are $(\fF\times\fB(L_k^4,x))$-measurable. Moreover, by assumption (\LVB, (ii)) (cf. \eqref{eq:LVB}, \eqref{eq:LVB.g}) the random variable\footnote{Recall that \emph{all} samples $V(\cdot, \th)$ are assumed to be smooth functions on $\Om=\DT^\nu$, so for every $\om'\in\Om$ and $x\in\DZ^d$ the value $V(T^x\om', \cdot)$ is a well-defined random variable on $\Th$.}
$gV(x;\om,\th)$, conditional on $\fF\times\fB(L_k^4,x)$,  does indeed admit a  probability density  bounded by
$\Cdprime   g^{-1} L_k^{B}$. As a result, for some $C_2<\infty$,
$$
\prmuj{ |\lam^{(j)}_{r} - E|\le 4 g \delta_k } \le C_2 \, L_k^{B} \delta_k
$$
and since the number of pairs $\big(\lam^{(j)}_{r}$, $\lam^{(1)}_{r}\big)$
is bounded by $(2R_j+1)^d (2R_1+1)^2 \le 9 L_k^{2d}$,
$$
\essup\, \prmu{ \cD_j  \,\big|\, \fBtj } \le C_3 \,  L_k^{2d+B} \delta_k \, .
$$
Next, one can re-write $\prmu{ \cap_{i\le j} \cD_i  }$ as follows:
\be\label{eq:proof.nuplustwo.R}
\bal
%
 \esmmu{ \esmmu{ \prod_{1\le i\le j} \one_{\cD_i} \,\big|\, \fBtj }  }
\le \mu \big\{ \bigcap_{1 \le i < j} \cD_i  \big\} \; \essup \prmu{ \cD_j \,\big|\, \fBtj }
\eal
\ee
(here $\esmmu{\cdot}$ is the expectation relative to $(\Th,\fB,\mu)$),
and by induction
$$
\prmu{ \cap_{i\le J_\nu} \cD_i  } \le C_4 \left(L_k^{2d+B}\right)^{J_\nu-1}
  \delta_k^{J_\nu-1}.
$$
The total number of families
$\{ \boxx_{R_j}(v^{(j)}) \subset\boxx_{L_k}(u^{(j)})\subset\boxxkfour(0)\}$ and arbitrary
$R_j\le L_k$ is bounded by
$\frac{1}{J_\nu!}L_k^{J_\nu (d + 1)}$, so that for $L_0$ large enough we obtain
$$
\prmu{ \cR(\om',k)  }
<  L_k^{(3d+B+1)J_\nu}  \delta_k^{J_\nu-1}.
$$
This completes the proof.
\qedhere

\begin{cor}\label{cor:small.var.centers}
Let $N_k\ge 1$ be an integer, $k\ge 0$; set $r_k = 1/(2N_k)$ and cover the torus $\DT^\nu$ by the cubes $\qube{r}{i}(\om_i)$, $1 \le i \le (N_k)^\nu =(2r_k)^{-\nu}$, with centers $\om_i$ of the form
$$
\om_i = ((2l_1+1) r_k, \, \ldots\, , (2l_\nu+1) r_k), \;\; l_1, \ldots, l_\nu \in[0, N_k-1]\cap \DZ.
$$
Using notations of Lemma \ref{lem:small.prob.qube}, introduce the event
\be\label{eq:def.cN.k}
\cN_k = \Th \setminus \bigcup_{1 \le i \le (N_k)^\nu} \cR(k,\om_i).
\ee
Then
\be\label{eq:cor.small.var.centers}
\bal
\prmu{ \Th\setminus \cN_k }
& \le  C_5(\nu,d ) \, L_k^{J_\nu(3d+B+1)} r_k^{ -\nu } \delta_k^{\nu +1}.  \\
\eal
\ee
\end{cor}

\proof
It suffices to apply Lemma \ref{lem:small.prob.qube} to each of the $(N_k)^\nu =(2r_k)^{-\nu}$ centers $\om_i$.
\qedhere
\smallskip

Now set
\be\label{eq:def.rk.deltak}
 r_k = \delta_k^{1+\frac{1}{2\nu}}, \;  k\ge 0,
\ee
and observe that
$$
r_k^{-\nu} \delta_k^{\nu + 1}
= \delta_k^{-\nu \cdot \frac{2\nu + 1}{2\nu} + \nu + 1} = \delta_k^{1/2},
$$
so that the bound \eqref{eq:cor.small.var.centers} takes the form
\be\label{eq:cor.small.var.centers.explicit}
\prmu{ \Th\setminus \cN_k }
 \le
C_5(\nu,d) \, L_k^{J_\nu(3d+B+1)} \delta_k^{1/2}.
\ee

\begin{lemma}
\label{lem:small.var.partition}
For all $\th\in\cN_k$, $E\in \DR$, $u\in\DZ^d$ and any $\om\in\Om$ there are at most $\nu+1$   pairwise disjoint lattice cubes $\boxx_{L_k}(u^{(j)})\subset\boxxkfour(u)$  which are $(E,\om,\th)$\emph{-PR}.
\end{lemma}

\proof
Since the gradient of the function $\om\mapsto v(\om; \th)$  is bounded (cf. \eqref{eq:gradient.bounded}), we have
$$
\forall\, \th\in\Th\;\; \sup_{\om\in \qube{r_k}{\om_i}} | gv(\om,\th) - gv(\om_i;\th)|
\le
C_6(\nu)\, g \, \diam(Q_{r_k}(\om_i)) \le C_6(\nu) g r_k
$$
and, therefore, for any $u\in\boxxkfour(0)$,
for $\delta_k$ small enough (so that $r_k =\delta_k^{1+\frac{1}{2\nu}}\ll \delta_k$)
\be\label{eq:sep.centers}
\sup_{\om\in \qube{r_k}{\om_i}}
\| H_{\boxx_{L_k}(u)}(\om,\th) - H_{\boxx_{L_k}(u)}(\om_i;\th) \|
\le C_6(\nu) g r_k \le \frac{1}{2} g \delta_k
\ee
By construction of $\cN_k$, for any $E\in \DR$,  if a cube $\boxx_{L_k}(u^{(1)})$ is $(E,\om,\th)$-R, then there are at most $\nu$ pairwise disjoint cubes  $\boxx_{L_k}(u^{(j)})$, $2 \le j \le \nu+1$, disjoint also with  $\boxx_{L_k}(u)$ and
such that if a cube $\boxx_{L_k}(w)$ is disjoint with the collection
$\{\boxx_{L_k}(u^{(j)}), 1\le j\le \nu+1\}$, then for any center $\om_i$, $1\le i \le N_k$,
and any cubes $\boxx_{R}(v)\subset\boxx_{L_k}(w)$,
$\boxx_{R_1}(v^{(1)})\subset\boxx_{L_k}(u^{(1)})$ with $R,R_1\in[L_{k-1},L_k]$
\be\label{eq:dist.spectra.1}
\dist( \Sigma(\boxx_{R}(v), \om_i;\th), \Sigma(\boxx_{R_1}(v^{(1)}), \om_i;\th))
\ge 4 g \delta_k \, .
\ee
Pick any $\om\in\Om$ and let $\qube{r_k}{i'}\subset\DT^\nu$, $i'=i'(\om)$,  be the cube
containing $\om$. Taking into account \eqref{eq:sep.centers}--\eqref{eq:dist.spectra.1}, we can write
$$
\ba
 \tabhigh{ \dist( \Sigomth{\om,\th}(\boxx_{R}(v)), \Sigomth{\om,\th}(\boxx_{R_1}(v^{(1)})) }\\
\quad \tabhigh{  \ge \dist( \Sigomth{\om_i,\th}(\boxx_{R}(v)), \Sigma(\boxx_{R_1}(v^{(1)}), \om_i;\th)) } \\
\qquad \tabhigh{ - \dist( \Sigomth{\om_i,\th}(\boxx_{R}(v)), \Sigomth{\om,\th}(\boxx_{R}(v)) }
  \tabhigh{ \,- \dist( \Sigomth{\om_i,\th}(\boxx_{R_1}(v^{(1)})), \Sigomth{\om,\th}(\boxx_{R_1}(v^{(1)}))) }
\\
\quad \tabhigh{ \ge 4 g \delta_k - 2\cdot \frac{1}{2} g \delta_k
>  2 g \delta_k }.
\ea
$$
Therefore, for any $ E\in \DR$ and any $\om\in\Om$
$$
\min \big\{ \dist( \Sigomth{\om,\th}(\boxx_{R}(v)), E),
\dist( \Sigomth{\om,\th}(\boxx_{R_1}(v^{(1)})), E) \big\} > g \delta_k.
$$
As a result, there is no collection of more than $\nu+1$ pairwise disjoint  $(E,\om,\th)$-PR cubes of radius $L_k$ with centers in $\boxxkfour(0)$.
\qedhere

\medskip\noindent
\textbf{Proof of Lemma \ref{lem:main.NMR}:}
It suffices to notice that, by Lemma \ref{lem:small.var.partition}, for every $\th\in \cN_k$ and any $\om\in\Om$, any cube $\boxx_{L_k}(u)$ is NMR (cf. Definition \ref{def:E.CNR}), and by Corollary  \ref{cor:small.var.centers} with the convention \eqref{eq:def.rk.deltak}, for $L_0$ large enough,
$$
\prmu{ \Th \setminus \cN_k }
\le  C_5(\nu,d) \, L_k^{J_\nu(3d+B+1)} \, \delta_k^{ 1/2 }
\le  g^{-a/2} e^{ -L_k^{b}}.
$$
\qed

\section{MSA for grand ensembles of deterministic operators}

\subsection{Initial scale bounds}

\begin{lemma}\label{lem:L0.NR.implies.NS}
Let $m>0$ and $\boxx_{L_0}(u)$ be an $(E,\om,\th)$\emph{-NR} cube, i.e.,
\be\label{eq:L0.NR}
\dist\left( \Sigma_{\om,\th}(\boxx_{L_0}(u)), E \right) \ge g \delta_0 = g^{1-a} e^{-4L_0^b}.
\ee
If  $g\delta_0 > 2d + 4de^{4\gamma(m,L_0)}$ and $L_0$ is large enough then  $\boxx_{L_0}(u)$ is $(E,m,\om,\th)$\emph{-NS}.
\end{lemma}

\proof

By min-max principle applied to the operators $gV(\om,\th)$ and
$H(\om,\th) = gV(\om,\th) + \Delta$, with $\| \Delta\|\le 2d$, the assumption of the Lemma implies that
$$
\dist\left( \Sigma_{\om,\th}(\boxx_{L_0}(u)), E \right) \ge g\delta_0 - 2d
\ge 4de^{4\gamma(m,L_0)} =: \eta >2.
$$
By Combes--Thomas estimate combined with \eqref{eq:L0.NR}, the Green functions obey
$$
|G_{}(x,y;E)|
\le \frac{2}{\eta} \exp\left( - \half \left(\ln \frac{\eta}{4d} \right) \|x-y\|_1\right)
< e^{-2\gamma(m,L_0) \|x - y \|},
$$
since $\eta>2$ and $\|x\|_1 \ge \max_i |x_i| = \|x\|$. For $L_0$ large enough, this implies \eqref{eq:def.NS}.
\qedhere

\subsection{Collections of singular cubes}
Define the subsets $\cT_k\subset\Th$ of the form
$$
\cT_k = \Big\{\th\,|\;
\pr{\om:\; \boxx_{L_k}(u) \text{ is $(\om,\th)$\rm{-T} } } \ge \delta_k^{1/2} \Big\}
$$
(recall that "$(\om,\th)$-T" stands for "$(\om,\th)$-tunneling", cf. Definition
\ref{def:E.CNR})
and
\be\label{eq:def.Th.k}
\Th_k = \bigcap_{l\le k} \left(\cN_l \setminus \cT_l\right) , \quad k=0, 1, \ldots,
\ee
Next, introduce the following statement, relative to the scale $L_k$, $k\ge 0$:
\par
\medskip\noindent
\Sparsek:
\emph{
For any $\th\in \Th_k$, all $\om\in\Om$, any $E\in \DR$ and any lattice cube $\boxxkfour(u)$
there exist at most $J_\nu - 1=\nu+1$ pairwise disjoint cubes $\boxx_{L_k}(u^{(j)})\subset\boxxkfour(u)$ which are $(E,m,\om,\th)$\emph{-S}.
}

\begin{remark}\label{rem:Sparsek.implies.good}
The property \Sparsek\, implies, in particular, that for any $(\om,\th)\in\Om\times\Th_k$,
any cube $\boxx_{L_k+1}(u)$ ($\subset\boxxkfour(u)$) must be $(m,\om,\th)$-good (cf. Definition\ref{def:E.m.NS}).
\end{remark}

\begin{remark}
The exponent $4$ in $L_k^4$ is quite arbitrary; replacing it by any larger  $D>0$ would not affect main arguments, but only modify technical constants. We chose the exponent $4>\alpha$ simply to stress that a cube of size much larger than $L_{k+1} = L_k^\alpha$ contains a limited number of simultaneously singular cubes of radius $L_k$. We believe that an optimal scale should be exponential or sub-exponential in $L_k$.
\end{remark}

\begin{lemma}\label{lem:sparse.L0}
The condition \Sparsez\, is fulfilled for sufficiently large $g$.
\end{lemma}

\proof
Consider an arbitrary cube $\boxx_{L_0^4}(u)$ and a number $E\in\DR$. By construction of the set $\Th_0$, for any $\om\in\Om$ there is a collection $\fS(E)$ of  at most $J_\nu-1$  cubes $\boxx_{L_0}(u^{(i)})\subset\boxx_{L_0^4}(u)$ such that any cube $\boxx_{L_0}(v)$ disjoint from $\fS(E)$  is $(E,m,\om,\th)$-NR. By Lemma \ref{lem:L0.NR.implies.NS}, if
$g^{1/2}\delta_0>2d+4de^{2\gamma(m,L_0)L_0}$, then such cubes $\boxx_{L_0}(v)$ must also be $(E,m,\om,\th)$-NS.
\qedhere

\begin{lemma}\label{lem:decay.GF.k}
If for some $\om\in\Om$, $\th\in\Th$ and $E\in\DR$ a cube $\boxx_{L_{k}}$ is $(m,\om,\th)$-good  and $(E,\om,\th)$\emph{-CNR}, then
it is also $(E,m,\om,\th)$\emph{-NS}.
\end{lemma}

\proof
The claim follows directly from Lemma \ref{lem:RDB.GF}; its idea goes back to \cite{FMSS85},\cite{DK89}.
\qedhere
\medskip

It is convenient to re-formulate Lemma \ref{lem:decay.GF.k} in the following way.

\begin{lemma}\label{lem:dicho.EF}
Assume the property \Sparsek, and let  $\th\in\Th_k$. If a cube $\boxxkone(u)$ is
$(E,\om,\th)$\emph{-CNR}, for some $E\in \DR$ and $\om\in\Om$,  then it is also $(E,m,\om,\th)$\emph{-NS}.
\end{lemma}

\proof
As was pointed out in Remark \ref{rem:Sparsek.implies.good},
if  $\th\in\Th_k$, then by assumed property \Sparsek\, any cube $\boxxkone(u)$ is
$(m,\om,\th)$-good.
Further,  $\boxxkone(u)$ is assumed to be $(E,\om,\th)$-CNR, so Lemma \ref{lem:decay.GF.k} implies that it is $(E,m,\om,\th)$-NS.
\qedhere

\subsection{Scale induction}

\begin{lemma}\label{lem:main.ind}
Statement \Sparsek\, implies \Sparsekone.
\end{lemma}

\proof
Pick any $\th\in\Th_{k+1} \subset \cN_{k+1}$ and any $E\in\DR$. By construction of the set $\cN_{k+1}$, for any $\om\in\Om$ there exists a collection $\fR_{k,u}(E,\om)$ of at most
$J_\nu - 1=\nu+1$  cubes $\boxx_{L_{k+1}}(u^{(j)}) \subset \boxx_{L_{k+1}^4}(u)$ such that any cube $\boxx_{L_{k+1}}(v)$ disjoint with $\fR_{k,u}(E,\om)$  must be $(E,\om,\th)$-CNR. Further, by assumption \Sparsek, for any $\om\in\Om$ the cube $\boxx_{L_{k+1}}(v)$ cannot contain $J_\nu$ or more disjoint $(E,m,\om,\th)$-S cubes of radius $L_k$, and by Lemma \ref{lem:decay.GF.k},  it must be $(E,m,\om,\th)$-NS. Therefore, any cube $\boxx_{L_{k+1}}(v)\subset\boxx_{L_{k+1}^4}(u)$ disjoint with $\fR_{k,u}(E,\om)$ is $(E,m,\om,\th)$-NS; this proves the assertion  \Sparsekone.
\qedhere

\medskip

Since the validity of \Sparsez\, is established in Lemma \ref{lem:sparse.L0}, we come, by induction, to the following conclusion:

\begin{theorem}\label{thm:sparse.all.k}
For $g$ large enough, the condition \Sparsek\, holds true for all $k\ge 0$.
\end{theorem}

\medskip

\subsection{Localization of eigenfunctions in finite cubes}

\begin{definition}\label{def:m.loc}
Given  a sample $v(\om,\th)$, a cube $\boxx_{L_{k}}(u)$ is called
$(m,\om,\th)$-localized (\thomloc) if for any eigenfunction $\psi_j$ of operator
$H_{\boxx_{L_{k}}(u)}(\om,\th)$ and any points $x,y\in\boxx_{L_{k}}(u)$ with
$\|x - y\| \ge L_k^{7/8}$
\be
|\psi_j(x) \, \psi_j(y)| \le e^{- \gamma(m,L_{k-1}) \|x - y\| },
\ee
otherwise, it is called $(m,\om,\th)$-non-localized (\thomnloc).
\end{definition}
Set
$
\Thinf = \Thinf(g) = \bigcap_{k\ge 0} \Th_k.
$
\begin{theorem}\label{thm:ind.prob}
\begin{enumerate}[\rm (A)]
  \item $\prmu{ \Thinf(g)} \ge 1 - C_7 g^{-a/2} $;
  \item for any $\th\in\Thinf$ and any $k\ge 0$,
\be
\pr{\om:\; \boxxk(u) \text{ is \thomnloc } } \le g^{-a/2} e^{- \, L_k^{b}}.
\ee
\end{enumerate}

\end{theorem}
\proof The first assertion follows directly from Corollary \ref{cor:small.var.centers} combined with assertion (B) of Lemma \ref{lem:Wegner} (cf. Eqn \eqref{eq:Wegner.3}). Further, let $\th\in\Thinf\subset\Th_k$. Then, by assertion (B) of Lemma \ref{lem:RDB.GF}, either $\boxxk(u)$ is $(\om,\th)$-tunneling, or it is \loc, so that, owing to Lemma \ref{lem:Wegner}, we have
$$
\bal
\pr{ \boxxk(u) \text{ is \thomnloc } }
& \le \pr{\boxxk(u) \text{ is $(\om,\th)$-T } }
\le g^{-a/2} e^{- L_{k-1}^{b}}.
\eal
$$
\qedhere

\subsection{Spectral localization: Proof of Theorem \ref{thm:main.SL}}

Fix any $\th\in\Thinf$ and let $\psi$ be a nontrivial, polynomially bounded solution of equation $H(\om,\th)\psi = E\psi$. There exists a point $\hu$ where $\psi(\hu)\ne 0$ and, as a result, there exists an integer $k_\circ$ such that for all $L\ge L_{k_\circ}$ the cube $\boxx_L(\hu)$ is $(E,m,\om,\th)$-S: otherwise, the $(E,m,\om,\th)$-NS property would imply, for arbitrarily large $L>0$,
$$
|\psi(\hu)| \le O(L^c) \,e^{-mL}  \tto{L\to\infty} 0.
$$
Let
$$
\Om'_j = \left\{
\om:\, \forall\, k\ge j\; \text{the cube $\boxx_{L_k}(\hu)$  is $(\om,\th)$-NT}
\right\}.
$$
Since $\th\in\Thinf$ and $\Thinf\cap\cT_k=\varnothing$, we have
$\pr{\om:\,\text{$\boxx_{L_k}(\hu)$  is $(\om,\th)$-T}} \le g^{-a/2} e^{-L_k^{b}}$, so it follows from  Borel-Cantelli lemma that for $\DP$-a.e. $\om$
there exists $k_1=k_1(\om)$ such that $\om\in\Om'_{k_1}$. Fix such an element $\om$ and set $k_2(\om) = \max\{ k_\circ, k_1(\om)\}$. From this point on, we will analyze the behavior of function $\psi$ at distances $\ge 3L_{k_2}$ from $\hu$.

Introduce annuli $\cA_k = \boxx_{3L_{k+1}}(\hu)\setminus\boxx_{3L_{k}}(\hu)$,
$k\ge k_2$, and let $x\in\cA_k$. Set
$R := \|x-\hu\| - 2L_{k-1} - 1$
and consider an arbitrary cube $\boxx_{L_{k-1}}(y)\subset\boxx_{R}(x)$.
Since $\boxx_{3L_{k+1}}(\hu)$ is $(\om,\th)$-NT and $3L_{k+1} < L_{k-1}^4$, either $\boxx_{L_{k-1}}(\hu)$ or $\boxx_{L_{k-1}}(y)$ must be $(E,\om,\th)$-CNR (cf. Definition \ref{def:E.CNR}).

Let us show that cube $\boxx_{L_{k-1}}(\hu)$ cannot be $(E,\om,\th)$-CNR. Indeed, since  $\th\in\Thinf$ by assumption, the cube $\boxx_{L_{k-1}}(\hu)$ contains less than $J_\nu$ disjoint $(E,m)$-S cubes of radius $L_{k-2}$. Combined with $(E,\om,\th)$-CNR property, this would imply, by virtue of Lemma \ref{lem:RDB.GF}, that $\boxx_{L_{k-1}}(\hu)$ is $(E,m)$-NS, which contradicts the choice of the scale $L_{k_\circ}$. Therefore, $\boxx_{L_{k-1}}(y)$ is $(E,\om,\th)$-CNR.

Using again  the assumption $\th\in\Thinf$, we see that every cube
$\boxx_{L_{k-1}}(y)\subset\boxx_{R}(x)$ contains less than $J_\nu$ disjoint $(E,m)$-S cubes of radius $L_{k-2}$ and is $(E,\om,\th)$-CNR. By Lemma \ref{lem:RDB.GF}, all  cubes $\boxx_{L_{k-1}}(y)$ are $(E,m)$-NS, so the same lemma implies that the cube $\boxx_{R}(x)$ itself is $(E,m)$-NS. Therefore, we can write, with the convention $\ln 0 = -\infty$,
$$
\qquad
\bal
\frac{\ln|\psi(x)|}{ \|x-\hu\|} & \le -\frac{\gamma(m,R)R}{\|x-\hu\| }
& \le - \frac{m(1 + \|x-\hu\|^{-1/8})  (\|x-\hu\|-3L_{k-1}) }{ \|x-\hu\|}
& \le -m.
\eal
$$
\qed

\section{Examples of regular grand ensembles}
\label{sec:GE.examples}

\subsection{"Randelette" expansions}

Following \cite{C01,C07}, consider a function $v:\Om\times\Th\to\DR$ given by a series of the form

\be\label{eq:def.randelettes}
v(\om,\th) = \sum_{n=0}^\infty a_n \sum_{k=1}^{K_n} \th_{n,k}\, \varphi_{n,k}(\om)
\ee
where the family of random variables $\{\th_{n,k}, n\in\DN, k\in[1,K_n]\}$ on $(\Th,\fB,\mu)$ is IID with bounded common probability density $\rho_\th$. Two particular choices are technically convenient: the  uniform distribution on $[0,1]$, where $\rho_\th(t) = \one_{[0,1]}(t)$, and the standard Gaussian distribution $\cN(0,1)$. Below we assume that $\rho_\th(t) = \one_{[0,1]}(t)$.

The functions $\ffi_{n,k}$ with the same value of $n$ (referred to as the $n$-th generation) are supposed to have a uniformly bounded overlap of their supports: for some $K'\in\DN$,
\be\label{eq:def.finite.overlap}
\sup_{n\ge 0} \; \sup_{\om\in\Om}\; \card\{k:\; \om\in \supp\, \ffi_{n,k} \} \le K'.
\ee
Following \cite{C01},  we will call these functions \emph{randelettes}, and the representation
\eqref{eq:def.randelettes} will be called a \emph{randelette expansion}.

In order to obtain samples $v(\cdot,\th)$ of class $\cC^M$, $M\ge 1$, the functions $\ffi_{n,k}$, have to be assumed of class $\cC^M(\Om)$. In the case of the uniform distribution, the random variables $\th_{n,k}$ (`siblings amplitudes') are bounded, so that the convergence of the series \eqref{eq:def.randelettes} is encoded in the decay properties of the `generation amplitudes' $a_n$, $n\ge 0$.

On the other hand, the random field of the form \eqref{eq:def.randelettes} has to fulfill the condition \LVB, and it is clear that an excessively rapid decay of amplitudes $a_n$ can destroy the local variation bound. We will show that for every smoothness class $\cC^M$, one can find an acceptable compromise between these two opposite requirements: convergence of the series \eqref{eq:def.randelettes} in $\cC^M$ and the `local freedom' condition \LVB.

\subsection{An example of $\cC^1$-randelettes on $\DT^1$}

The general structure of randelette expansions, as well as the term "randelette", is clearly inspired by wavelets (\emph{ondelettes}, in French). However, the orthogonality issues are of little importance here, and the finite-overlap condition, serving as a substitute of orthogonality, is more than sufficient for applications to Wegner-type estimates and to localization theory.

Consider the following function on $\DR$:
$$
\ba
\phi(t) = \frac{t^2}{2} \one_{[0,1)}(t)
+ \big(1 - \frac{(t-2)^2}{2}\big) \one_{[1,2)}(t) + \one_{[2,+\infty)}(t).
\ea
$$
By direct inspection, one can check that $\phi\in \cC^1(\DR)$ and $\|\phi\|_{\cC^1(\DR)}=1$. Similarly, the function $t\mapsto\phi(12 - t)$ has unit $\cC^1(\DR)$-norm, and so does the product
$\Phi(t) = \phi(t)\phi(12-t)$, which vanish outside the interval $(0,2^4)$ and equals $1$ on the segment $[2,14]$. Further, define a sequence of scaled functions
$$
\Phi_n(t) = \Phi(2^{n} t), \quad \|\Phi_n\|_{\cC^1(\DR)} = 2^{n},
$$
with $\supp \Phi_n = [0, 2^{4-n}]$, and their shifts
$$
\ffi_{n,k}(t) = \Phi_n(t-k), \quad k\in\DZ,
$$
with $\supp \ffi_{n,k} = \left[\frac{k}{2^{n-4}}, \frac{k+1}{2^{n-4}}\right]$. Using the natural projection $\DR \to \DR/\DZ = \DT^1$, one can consider $\ffi_{n,k}$ as functions on the unit circle $\DT^1$, and it is clear that
\begin{itemize}
  \item the family $\{\phi_{n,k}, 1 \le k \le 2^n\}$ has a bounded overlap ($K_n\le 2^4$),
  \item each point $t\in\DT^1$ is covered by a segment  on which at least one of the functions $\ffi_{n,k}$ identically equals $1$.
\end{itemize}

Form now a randelette expansion  \eqref{eq:def.randelettes} with $a_n = e^{cn}$. Since $\|\ffi_{n,k}\|_{\cC^1} = O(2^n)$, we see that for $c$ large enough the series \eqref{eq:def.randelettes} converges uniformly in $\cC^1(\DT^1)$, regardless of the values of the random coefficients $\th_{n,k}$ (all of them are bounded by $1$).  So, we obtain an example of a smooth randelette expansion on the one-dimensional torus $\DT^1$.

\subsection{Randelettes of class $\cC^M$}

An adaptation of the previously described construction to the case where the series \eqref{eq:def.randelettes} is to be of class  $\cC^M$  is fairly straightforward. Indeed, for any $M\ge 1$ there exists a $\cC^M$-function $\Phi:\DR\to\DR$ equal to $1$ on an interval $[1/4,3/4]$ and vanishing outside $[0,1]$; it can be easily constructed explicitly, e.g., with the help of the so-called $B$-splines (convolutions of indicator functions of finite intervals). Then for the functions $\ffi_{n,k}(t) = \Phi(2^n(t-k))$ we have
$$
\|\ffi_{n,k}\|_{\cC^M(\DR)} \le 2^{Mn} \|\Phi\|_{\cC^M(\DR)} \le \Const e^{-c'Mn}.
$$
Therefore, for $c>0$ large enough and $a_n = e^{-cn}$ the randelette expansion
\eqref{eq:def.randelettes} converges uniformly in $\cC^M(\DT^1)$, regardless of the values of the coefficients $\th_{n,k}$ which are bounded by $1$.

A similar construction can be extended to the torus $\DT^\nu$, $\nu\ge 1$, by taking the "mother" randelette as the tensor product of its one-dimensional counterparts:
$$
\Phi(t_1, \ldots, t_\nu) = \Phi(t_1) \cdots \Phi(t_\nu)
$$
and then defining scaled and translated functions $\ffi_{n,k}(t_1, \ldots, t_\nu)$. Again, the randelette expansion with functions $\ffi_{n,k}\in\cC^M(\DT^\nu)$  converges in
$\cC^M(\DT^\nu)$, when the generation amplitudes have the form $a_n = e^{-cn}$, with sufficiently large $c>0$ proportional to $M$.

\subsection{Validity of the Local Variation Bound}

Let us show that the randelette expansions of an arbitrary smoothness class $\cC^M$, with amplitudes $a_n=e^{-cn}$ and arbitrarily large $c>0$ satisfy \LVB\, for some $B=B(M)\in(0,+\infty)$. This is the central point of our construction, allowing to apply the MSA approach to deterministic operators and to replace a complicated differential-geometric analysis of "small denominators", appearing in the course of scaling procedure, by simpler probabilistic arguments.

Given a positive integer $N$, the series \eqref{eq:def.randelettes} can be re-written as follows:
$$
\bal
v(\om,\th) &=  \sum_{n=0}^{N-1} a_n \sum_{k=1}^{K_n} \th_{n,k}\, \varphi_{n,k}(\om)
+  \sum_{n=N}^\infty a_n \sum_{k=1}^{K_n} \th_{n,k}\, \varphi_{n,k}(\om) \\
& =  S_N(\om,\th) + \sum_{n=N}^\infty a_n \sum_{k=1}^{K_n} \th_{n,k}\, \varphi_{n,k}(\om)
\eal
$$
where $S_N$, considered as a random variable on $\Th$, is measurable with respect to the sigma-algebra $\fB_N$ generated by random coefficients $\{ \th_{n,k}, n<N, 1\le k \le K_n\}$, while the remaining series in the RHS is $\fB_N$-independent (again, as  a random variable on $\Th$). Introduce the sigma-algebra $\fB_N^\Om = \fF\times \fB_N$. Conditional on $\fB_N^\Om$, all functions
$(\om,\th)\mapsto\ffi_{n,k}(\om)$ become non-random, as well as the functions
$(\om,\th)\mapsto\th_{n,k}$ with $n< N$, while the random variables $(\om,\th)\mapsto\th_{n,k}$
with $n\ge N$ are  $\fB_N^\Om$-independent.

Fix an integer $L>1$ and points $u\in\DZ^d$, $x,y\in\boxx_L(u)$ with $x\ne y$. Since
$\|x-y\|\le 2L$, the condition \USR\, implies that for any $\om\in\Om$
$$
\dist(T^x\om, T^y\om) \ge C (2L)^{-A}.
$$
As a result, the points $T^x\om, T^y\om$ are separated by the supports of all functions $\ffi_{n,k}$ with $n\ge N$, provided that
$$
\max_{n\le N}\; \max_{k} \; \diam\, \supp\, \ffi_{n,k} < \half C (2L)^{-A}.
$$
By construction of the functions $\ffi_{n,k}$, we have
$\diam\, \supp\, \ffi_{n,k} \le 2^{4-n}$, so that the above requirement is fulfilled for
\be\label{eq:sep.N.L}
\bal
 N & \ge N(L, A, C) := a(A,C) \ln L + b(A,C)
\eal
\ee
with $a(A,C) = \frac{A}{\ln 2}$, $b(A,C) =  A+5 - \frac{\ln C}{\ln 2}$. We see that for $N$ starting from $[N(L,A,C)]+1 = O(\ln L)$, no pair of phase points $T^x\om, T^y\om$ with $x,y\in\boxx_L(u)$ can be covered by the support of the same function $\ffi_{n,k}$ with $n\ge N$. Therefore, no random variable  $\th_{n,k}$ with $n\ge N$ can affect two distinct values of the random potential $gV(x;\om,\th)$, $gV(y;\om,\th)$ in any cube of radius $L$. Conditional on $\fB_N^\Om$, all values $\{gV(x;\om,\th), x\in\boxx_L(u)\}$ become (conditionally!) independent.
Conditioning further on all $V(y;\om,\th)$ with $y\ne x$ does not change the conditional distribution of $V(x;\om,\th)$, so it suffices to examine the conditional probability density
of $V(x;\om,\th)$ given $\fB_N^\Om$.

The latter does exist, since, by construction of the randelette expansion, for every $n\ge 1$, every point  of the torus, including $T^x\om$, is covered by an interval where some function $\ffi_{n,k}$ with $k = k(x,n,\om)$, equals $1$. Therefore,
$$
V(x;\om,\th) = v(T^x\om,\th) = S_N(\om,\th)+ a_n\th_{n, k(x,n,\om)} \cdot 1 + \xi(\om,\th),
$$
where $S_N(\om,\th)$ is $\fB_N^\Om$-measurable and $\xi$ is a sum of random variables conditionally independent of $\th_{n,k(x,n,\om)}$. Since $\xi$ is (conditionally) independent of
$\th_{n, k(x,n,\om)}$, their sum admits a probability density given by the convolution of the probability density of $\th_{n, k(x,n,\om)}$ with the probability distribution of $\xi$; this operation does not increase the sup-norm of the density.

Finally, the random variable $a_n\th_{n, k(x,n,\om)}$ is uniformly distributed in $[0, a_n]$, so that its probability density is bounded by $a_n^{-1} \le e^{cn}$. Setting $n=N(L,A,C) = a\ln L + b$, we get an upper bound on the conditional probability density
of $V(x;\om,\th)$ of the form
$$
\| p_{x}(\cdot | \fB_N^\Om) \|_\infty \le e^{\Const \, \ln L} \le L^B,
\quad B = B(A,C)\in(0,+\infty).
$$

\section{On Minami-type bounds for generic deterministic operators}
\label{sec:Minami}

As was shown in \cite{GV07}, \cite{BHS07}, the spectral spacings $|E_j - E_i|$ of a random LSO in a cube $\boxx_L(u)$ are positive with probability one, provided that the random potential $V(x;\om)$ is an IID random field with bounded marginal probability density $\rho$. Specifically, for any bounded interval $I\subset\DR$,
$$
\pr{ \tr \Pi_I\big( H_{\boxx_L(u)}(\om) \big) \ge J} \le \frac{( \pi \|\rho\|_\infty)^J}{J!} |I|^J.
$$
A direct inspection of the proofs evidences that the requirement of independence of the potential field can be substantially relaxed: given an integer, $J\ge 2$, it suffices that, for any collection of $J$ pairwise distinct points $\cX_J = \{x_1, \ldots, x_J\}$,  the joint conditional probability distribution of the random variables $V(x_1;\om), \ldots, V(x_J;\om)$ admit a bounded conditional probability density, given all values $\{V(y;\om), y \in \boxx_L(u) \setminus \cX_J \}$. In other words, main results of \cite{GV07}, \cite{BHS07} can be re-formulated in the following way.

\begin{prop}[Cf. \cite{GV07}, \cite{BHS07}]\label{thm:Mi.IID}
Assume that the random field $V:\DZ^d\times\Om\to\DR$ fulfills the following condition:
for any cube $\boxx_{L}(u)$ and any subset $\cX_J = \{x_1, \ldots, x_J\}$, $\card \cX_J=J$,
with fixed $J\ge 1$, the joint conditional probability distribution of the vector $(V(x_1;\om, \cdots, V(x_J;\om))$, given $\{V(y;\om), y \in \boxx_L(u) \setminus \cX_J \}$, admits a bounded probability density $\rho(t_1, \ldots, t_J)\le C_\rho$. Then for any bounded interval $I\subset\DR$,
$$
\pr{ \tr \Pi_I\big( H_{\boxx_L(u)}(\om) \big) \ge J} \le \frac{( \pi C_\rho)^J}{J!} |I|^J.
$$
\end{prop}

\begin{theorem}\label{thm:GE}
Consider a regular randelette expansion of the form \eqref{eq:def.randelettes}. For sufficiently large $B'\in(0,+\infty)$, any finite interval $I\subset\DR$, any $L>0$ and some $C_{10}\in(0,+\infty)$
$$
\prpmu{ \tr \Pi_I\big( H_{\boxx_L(u)}(\om,\th) \big) \ge J }
\le C_{10} L^{B'} |I|^J.
$$
\end{theorem}
\proof
Re-write \eqref{eq:def.randelettes} as follows:
$$
\bal
v(\om,\th)
 =  S_N(\om,\th) + \sum_{n=N}^\infty a_n \sum_{k=1}^{K_n} \th_{n,k}\, \varphi_{n,k}(\om)
\eal
$$
and set $N = N(L, A, C) = a(A,C) \ln L + b(A,C)$ with $a(A,C) = \frac{A}{\ln 2}$,
$b(A,C = A + 5 - \frac{\ln C}{\ln 2}$. Then any pair of phase points $T^x\om, T^y\om$ with $x\ne y$ is separated by the supports of  functions $\{\ffi_{n,k}, 1 \le k \le K(n)\}$ with any $n\ge N(L,A,C)$. Therefore, conditional on all $\th_{n,k}$ with $n \le N(L,A,C)-1$ and with $n \ge N(L,A,C)+1$, the values of the potential $V(T^x;\om,\th)$ at different points $x\in\boxx_L(u)$ become independent and have uniform distributions in (different) intervals of length $a_N^{-1}$. Now the claim follows directly from Proposition \ref{thm:Mi.IID}.
\qedhere

\section*{Appendix}

\subsection{Proof of Lemma \ref{lem:Wegner}}

\textbf{(A)} Fix a cube $\boxx_{L_k}(u)$.
We will seek first a bound for the LHS of \eqref{eq:Wegner.1} conditional on sigma-algebra $\fB_\Om (\cong \fF)$ generated by random variables $(\om,\th) \mapsto \om$ on $\Om\times\Th$. By assumption \USR, if $\om\in\Om$ and  $x\in\boxx_{L_k}(u)$ are fixed, all phase points
$\{T^y\om, x \ne y\in\boxx_{L_k}(u)$ lie outside the cube $Q_{L^{-A}}(\om)\subset\Om$.

Furthermore, conditional on $\om$ and on $\fB(L_k,x)$, all values of the potential $gV(y;\om,\th)$ with $y\in\boxx_{L_k}(u)\setminus\{x\}$ become non-random (measurable with respect to the condition), while the remaining random value $gV(x;\om,\th)$ admits a conditional probability density bounded by $\Cdprime  L_k^B g^{-1}$, owing to assumption \LVB. Applying the conventional Wegner bound (see, e.g., \cite{CL90})  to this conditional measure, we can write
$$
\prpmu{(\om,\th):\,   \dist( \Sigma_{\om,\th}(\boxx_{L_k}(u)), E) \le s
\, | \, \fB_\Om, \fB(L_k,x) }  \le C_8\, \cdot L_k^d\, \Cdprime   L_k^{B}g^{-1} \, s
$$
yielding
\be\label{eq:prob.1.R.box.proof.1}
\ba
\diy \prpmu{(\om,\th):\; \;   \dist( \Sigma_{\om,\th}(\boxx_{L_k}(u)), E) \le s} \\
\tabhigh{\diy \quad = \esmommu{\prpmu{(\om,\th):\; \;
           \dist( \Sigma_{\om,\th}(\boxx_{L_k}(u)), E)  \le s
\, | \, \fB_\Om, \fB(L_k,x) }    } }
\\
\diy \tabhigh{ \quad \le C_9\, \cdot L_k^d\, L_k^{B} g^{-1}\, s}\, .
\ea
\ee

\noindent
\textbf{(B)}
Consider disjoint cubes $\boxx_{\ell}(x')$, $\boxx_{\ell}(x'')\subset\boxx_{L_{k}}(u)$,
$\ell\ge L_{k-1}$, and let $x\in\boxx_{L_{k-1}}(x')$. Observe that conditioning on
$\fB(L_k, x)$, used in the previous argument, fixes not only the values $gV(y;\om,\th)$ with $y\in\boxx_{\ell}(x')\setminus\{x\}$, but  all values with  $y\in\boxx_{L_k}(u)\setminus\{x\}$. This includes the sample of the  potential in the cube $\boxx_{\ell}(x'')$ disjoint from $\boxx_{\ell}(x')$. Therefore,  conditional on $\fB(L_k,x)$, the spectrum  $\Sigma_{\om,\th}(\boxx_{\ell}(x''))$ also becomes non-random. If $\boxx_{\ell}(x')$ and $\boxx_{\ell}(x'')$ are $(E,\om,\th)$-R for some $E$, then
$$
\dist( \Sigma_{\om,\th}(\boxx_{\ell}(x')), \Sigma_{\om,\th}(\boxx_{\ell}(x'')))
\le 2 g \delta_k = 2 g^{1-a} e^{-4L_{k-1}^{b}}.
$$
By assertion (A), for each of $(2\ell+1)^d$ eigenvalues $E''_j$ of operator $H_{\boxx_{\ell}(x'')}(\om,\th)$ we can write
\be\label{eq:prob.1.R.box.proof.2}
\prpmu{(\om,\th):\; \;   \dist( \Sigma_{\om,\th}(\boxx_{\ell}(x')), E''_j) \le 2g\delta_k}
 \le C_9\,  L_k^{d+B} \,  g^{-1} \cdot\, g\delta_k.
\ee
Now the bound \eqref{eq:Wegner.2} follows from
\eqref{eq:prob.1.R.box.proof.2} by Chebyshev's inequality, since the number of all pairs $x',x''\in\boxx_{L_k}(u)$ is bounded by $(2L_k+1)^{2d}/2$, and $\ell\le L_k$ takes less than $L_k$ possible values.
\qed

\subsection{"Radial descent" bounds}

In \cite{C08}, we introduced the following notion.

\begin{definition}
Consider a set $\Lam\subset\DZ^d$
and a bounded function $f:\,\Lam\to\DC$. Let $\ell\ge 1$ be an integer and $q>0$. Function $f$ is called $(\ell,q)$-subharmonic in $\Lam$ if for any $u$ with $\dist(u, \pt \Lam)\ge \ell$, we have
\be\label{eq:subh.1}
|f(u)| \leq q \;\;\mymax{y:\, \|y-u\| \le \ell+1} |f(y)|
\ee

\medskip
\noindent
Function $f$ is called $(\ell,q, \cS)$-subharmonic, with $\cS\subset \Lam$, if for any
$u\in \cR :=\Lam \setminus \cS$ the bound \eqref{eq:subh.1} holds,
while for any $x\in\cS$ with $\dist(x, \pt \Lam)\ge \ell$
\be\label{eq:def.subh.2}
 |f(x)| \leq q \;\;\mymax{y:\, \|x - y\| \le r(x)+\ell} |f(y)|,
\ee
where
\be\label{eq:def.r.x}
r(x) = \min\{ r\ge \l+1:\,  \Lam_{r+\ell}(x) \setminus \Lam_{r-\ell}(x) \subset \cR \},
\ee
provided that the set of values $r$ in the RHS is non-empty. In all other cases, no specific upper bound on $|f(x)|$ is assumed.
\end{definition}

\begin{lemma}[Cf. \cite{C08}]\label{lem:SubH} Let $f$ be an $(\ell,q,\cS)$-subharmonic function on
$\boxx_L(u)$.  Suppose that $\cS$ can be covered by a collection of cubes $Q_1, \dots, Q_K$ with $\sum_i \diam Q_i \le \cW$. Then
$$
 |f(u)| \leq q^{[(L-\cW)/\ell]} \cM(f,\boxx).
$$
\end{lemma}

The motivation for the above definition comes from the following observations.
Consider a pair of cubes  $\boxx_\ell(u) \subset \boxx_L(x_0)$. If $\boxx_\ell(u)$ is $(E,m)$-NS, then the Geometric Resolvent Identity,
$$
G_{\boxx_{L}(x_0)} (u,y)
= \sum_{(w,w')\in\pt \boxx_\ell(u)} G_{\boxx_{\ell}(u)} (u,w) \,
G_{\boxx_{L}(x_0)} (w',y), \quad y\not \in \boxx_{\ell}(u),
$$
implies that function $f: x \mapsto G_{\boxx_L(x_0)}(x,y;E)$ satisfies, with
$
q =  e^{-\gamma(m,\ell)\ell},
$
$$
|f(u)|  \le  q  \cdot \mymax{v:\, \|v-u\|=\ell} |f(v)|.
$$
Consider a more general case where $\boxx_L(x_0)$ contains at most $K$  cubes, $\{\boxx_\ell(u_j), 1\le j\le K\}=:\fS$, such that any cube $\boxx_\ell(v)$ disjoint  with $\fS$ is $(E,m)$-NS. Define $\cS$ as the $(\ell+1)$-neighborhood of $\fS$ in $\boxx_L(x_0)$,
$\cR :=\boxx_L(x_0)\setminus \cS$, and let $r(x)$ be defined for $x\in\cS$ as in \eqref{eq:def.r.x}. It is clear that $\cS$ can be covered by a collection of cubes
$\boxx_{r_i}(u_i)$ with $W:=\sum_i 2r_i\le K(2\ell+1)+\ell+1$. Assume that $\boxx_L(x_0)$ is $E$-CNR (hence, any cube $\boxx_R(x)$ with $R\ge \ell$ is $E$-R), and pick $x\in\cS$.
Applying the GRI twice, we get
$$
\bal
|f(x)| &\le O(\ell^{d-1})\, e^{4L^{b}} \max_{\|w-x\|\le r(x)} | f(w) | \\
& \le  e^{-m\ell - m\ell^{7/8} + 4\ell^{3/8} +O(\ln \ell)} \max_{\|v-x\|\le r(x)+\ell+1} | f(v) |
\le e^{ -m\left(\ell + \frac{1}{2}\ell^{7/8}\right)}
\eal
$$
if $\ell$ is large enough. In other words, $f$ is $(\ell,q,\cS)$-subharmonic with
$q=e^{ -m(\ell + \frac{1}{2}\ell^{7/8})}$. Similarly one can treat the eigenfunction correlators $f:(x,y)\mapsto \psi_j(x)\psi_j(y)$.

A direct application of Lemma \ref{lem:SubH} leads to the following statement which can be considered as a variant of a well-known technical result going back to papers \cite{FMSS85}, \cite{DK89}.

\begin{lemma}[Cf. \cite{C08}]\label{lem:RDB.GF}
Fix an integer $K\ge 1$ and suppose that for any $E\in\DR$ a cube $\boxx_{L_k}(u)$
contains no collection of $K$ pairwise disjoint $(E,m)$\emph{-S} cubes of radius $L_{k-1}$.
\begin{enumerate}[\rm (A)]
  \item If for some $E\in\DR$, the cube $\boxx_{L_k}(u)$ is $(E,\om,\th)$\emph{-CNR}  then for $L_0$ large enough,  it is $(E,m)$\emph{-NS}.
  \item If the cube $\boxx_{L_k}(u)$ is $(\om,\th)$\emph{-NT}, then for $L_0$ large enough,  it is $m$-localized.
\end{enumerate}

\end{lemma}

\proof
The assertion (A) follows directly from Lemma \ref{lem:SubH} applied to the function
$f:x\mapsto G_{\boxx_{L_k}(u)}(u,x;E)$. To prove (B), we make use of the subharmonicity of the functions $(x_1,x_2)\mapsto \psi_j(x_1)\psi_j(x_2)$ both in $x_1$ and $x_2$; here $\psi_j$ is an eigenfunction with eigenvalue $E_j$. More precisely, $(\om,\th)$-NT property allows, for any $E_j$,  to exclude some cube $\boxx_{L_{k-1}}(w)$ such that any cube $\boxx_{L_{k-1}}(v)$ disjoint with it is $(E_i,\om,\th)$-NR. With $x_1,x_2$ fixed, set $r_i = \min(0, \|x_i - w\|-2 L_{k-1}-2)$, $i=1,2$, so that
$r_1+r_2\ge L_k^{7/8} - O( L_{k-1})$. If $r_i>0$, then no cube
$\boxx_{L_{k-1}}(y)\subset\boxx_{r_i+ L_{k-1}}(x_1)$ is $(E,\om,\th)$-R, and one can apply Lemma \ref{lem:SubH} to the function
$f:x_i \mapsto \psi_j(x_i)\psi_j(x_{2-i})$ and obtain
$$
|\psi_j(x_i)|\le e^{-m\left(1 + \frac{1}{2} L_{k-1}^{-1/8}\right)(r_i - O(L_{k-1})}.
$$
If both $r_1>0$ and $r_2>0$, then (B) follows by a straightforward calculation from
$$
|\psi_j(x_1) \psi_j(x_2)|
\le e^{-m\left(1 + \frac{1}{2} L_{k-1}^{-1/8}\right) \left(r_2+r_2- O(L_{k-1}) \right)},
$$
taking into account that $r_1+r_2\ge L_k^{7/8} - O( L_{k-1})$, and if one of the radii $r_i$, $i=1,2$, is zero, the claim follows from the bound on the remaining value  $|\psi_j(x_{2-i})|$.
\qedhere

\section*{Acknowledgements}

It is a pleasure to thank Ya. G. Sinai, Tom Spencer, Misha Goldstein and Lana Jitomirskaya for fruitful discussions of localization techniques for deterministic random operators.


\end{document}